# A Comparison with two semantic sensor data storages in total data transmission

Manaf Sharifzadeh[1]
Department of Computer,Firoozkooh Branch,Islamic Azad University, Firoozkooh, IR, Email : manafsharif@gmail.com

saeid aragy[2]
department of computer, Science and research branch, Islamic Azad University,Khomein, Iran, Email : saeed.aragy@gmail.com

Kaveh Bashash[3]
Department of Computer,Damavand Branch, Islamic Aazad University, Damavand, Iran, Email : kvb2002@yahoo.com

Shahram Bashokian[4]
School of Computer Engineering,Iran University of Science and Technology(IUST), Email : Bashokian@comp.iust.ac.ir

,mehdi gheisari[5]
* varamin university of science and technology, Email : mehdi.gheisari61@gmail.com

*Abstract*—The creation of small and cheap sensors promoted the emergence of large scale sensor networks. Sensor networks allow monitoring a variety of physical phenomena, like weather conditions (temperature, humidity, atmospheric pressure ...), traffic levels on highways or rooms occupancy in public buildings. Some of the sensors produce large volume of data such as weather temperature. These data should be stored somewhere for user queries. In this paper two known sensor data storage methods that store data semantically has been compared and it has been shown that storing data in ontology form consumes more energy so the lifetime of sensor network would decreases. The reason we choose them is that they are useful and popular.

*Keywords*- wireless, semsos, SWE

## I. INTRODUCTION

Progresses in wireless communications and micro electromechanical systems (MEMS) led to the deployment of large-scale *wireless sensor networks* (WSN), in other words it revolutionized the way we monitor and control environments of interest [1, 2]. WSN was identified as one of the ten emerging technologies that will change the world in MIT Technology Review [3]. A wide variety of attractive applications with the use of WSN [4] would come into reality, such as habitat monitoring, [5] search and military industries, disaster relief, target tracking, precision agriculture and smart environments.WSN creates variant types of data like arrays and images. These data should be stored somewhere for variety of queries. The paper exemplifies how the use of semantics can enhance data management in sensor networks. Semantics exploit underlying relationships between data captured by sensors [6-8].Section 2 describes some background knowledge like XML and RDFa. Section 3 describes SSW framework. Section 4 describes a semantic data storage. Section5provides an evaluation of the work. Finally in Section6we discuss our conclusions.

## II. Background Knowledge

This section describes some background knowledge we should have.

A. XML (Extensible Markup Language)

XML is the abbreviation of Extensible Markup Language.XML includes a set of rules for defining semantic tags that break a document into different parts and defines those different parts of the document [6].
XML is a meta-markup language that defines a syntax in which other domain-specific markup languages can be written. Syntactically, XML documents look like HTML documents. A well-formed XML document—one that conforms to the XML syntax—contains exactly one element. Additionally an arbitrary number of comments and processing instructions can be included.
XML introduces some languages to allow more semantic management of information than HTML. XML is about the description of data, with nothing said about its presentation.HTML combines some fundamental descriptive markup, plus a great deal of mark up that describes the presentation of the data [7].

B. Sensor Web Enablement (SWE)

The Open Geospatial Consortium recently built the Sensor Web Enablement as a suite of specifications related to sensors, sensor data models, and sensor web services that would permit sensors to be accessible and controllable through the Web [8,9].

The core language and service interface includes the following:
(1) Observations & Measurements (O&M) - Standard models and XML Schema for encoding observations and measurements from a sensor, both archived and real-time.
(2) Sensor Model Language (Sensor ML) - Standard models and XML Schema for describing sensors systems; in other words it provides information needed for discovery of sensors, location of sensor observations.
(3) Transducer Model Language (Transducer ML) – Standard models and XML Schema for supporting real-time streaming of data to and from sensor systems.
(4) Sensor Observations Service (SOS) - Standard web service for requesting, filtering, and retrieving observations and sensor system information. This is the intermediary between a client and an observation source or near real-time sensor channel.

The following example shows a timestamp encoded in O&M and semantically annotated with RDFa.
The timestamp's semantic annotation describes an instance of time: Instant (here, time is the namespace for OWL-Time ontology):

<swe: component rdfa: about="time_1"
rdfa: instance of ="time: Instant">
<swe: Timerdfa: property="xs: date-time">
2010-0308T05:00:00
</swe: Time>

</swe: component>

This example generates two RDF triples. The first, time_1 rdf: type time: Instant, describes time_1 as an instance of time: Instant (subject is time_1, predicate is rdf: type, object is time: Instant). The second, time_1 xs: date-time "2010-03-08T05:00:00,"describes a data-type property of time_1 specifying the time as a literal value (subject is time_1, predicate is xs: date-time, object is "2008-03-08T05:00:00")[10].

### C. RDFa (or Resource Description Framework - in - attributes):

Many languages can be used for annotating sensor data, such as RDFa, XLink, and SAWSDL (Semantic Annotations for WSDL and XML Schema).

Here, we describe the use of RDFa, a W3C proposed standard (www.w3.org/2006/07/SWD/RDFa/) and a markup language that enables the layering of RDF information on any XHTML or XML document. RDFa is a set of extensions to XHTML. RDFa uses attributes from XHTML's meta and link elements and generalizes them so that they are usable on all elements. This allows annotating XHTML markup with semantics RDFa provides a set of attributes that can represent semantic metadata within an XML language from which we can extract RDF triples using a simple mapping[11].

### III. SSW (semantic sensor Web)

Seth and Hanson [10] discuss the idea of a semantic sensor Web framework. SSW is used for providing enhanced meaning for sensor observations so as to enable situation awareness. It enhances meaning by adding semantic annotations to existing standard sensor languages of the SWE. These annotations provide more meaningful descriptions and enhance access to sensor data than SWE alone, and they act as a linking mechanism to bridge the gap between the primarily syntactic XML-based metadata standards of the SWE and the RDF/OWL-based metadata standards of the Semantic Web. In association with semantic annotation, ontologies and rules play an important role in SSW for interoperability, analysis, and reasoning over heterogeneous multimodal sensor data.

### IV. ES3N

ES3n uses Semantic Web techniques to manage and query data collected from a mini dome Sensor Network. Our tool supports complex queries on both continuous and archival data, by capturing important associations among data, collected and stored in a distributed dynamic ontology [12].

### V. Implementation, Evaluation and Comparison

At first we have evaluated these two methods using j-sim[13,14] software that is a sensor network simulator in 10. For

evaluation of SSW we use the following data:

```
<swe: Data Record definition="urn: ogc: def: property: OGC: atmospheric Conditions>
<swe: fieldswe-om: Quantityrdf: about="#AirTemperature" name=AirTemperature">
<swe: quantitiy definition=" urn: ogc: def: property: OGC: AirTemperature">
<swe: uom code="Cel" swe-om: hasUomIdentifierrdf:about= "http://sweet.jpl.nasa.gov/ontology/units.owl#degreeC"/>
<swe:valueswe-om:hasDoubleValuerdf:daatype="&xsd;double">35.1</swe:value>
</swe: Quantitiy>
</swe: field>
<swe: fieldswe-om: Quantityrdf: about="#AirTemperature" name=Winspeed">
<swe: quantitiy definition=" urn: ogc: def: property: OGC: WinSpeed">
<swe: uomswe-om: hasUomIdentifierrdf: about= "http://sweet.jpl.nasa.gov/ontology/units.owl#meter_persecond" code="m/s"/>
<swe:valueswe-om:hasDoubleValuerdf:daatype="&xsd;double">6.5</swe:value>
</swe: Quantitiy>
</swe: field>
</swe: DataRecord>
```

This example generates two RDF. The first air temperature is 35.1 Celsius that data type is double and the next shows wind speed is 6.5 meter per second.

After that we show above example in ontology form:

```
<swe-om: Quantity rdf:Id="Quantity_AirTemperature">
<swe-om: hasUomIdentifier rdf:Resource="http://sweet.jpl.nasa.gov/ontology/units.owl#degreeC"/>
<swe-om: has DoubleValuerdf: DataType=http://www.w3.org/2001/XML.Schema# Double>35.1</ swe-om: has DoubleValue>
<swe-om: has Namexml:lang="en">air temperature</ swe-om:hasName>
<swe: has Definition rdf:datatype=http://www.w3.org/2001/XML.Schema#anyURI>urn:ogc:def:property:OGC:AIRTemperature</swe:hasDefinition>
</swe-om: Quantity>
<swe-om: Quantity rdf:Id="Quantity_WinSpeed">
<swe: has Definition rdf:datatype=http://www.w3.org/2001/XML.Schema#anyURI>urn:ogc:def:property:OGC:WinSpeed</swe:hasDefinition>
<swe-om: has Namexml:lang="en">Win Speed</ swe-om:hasName>
<swe-om: has UomIdentifier rdf:Resource="http://sweet.jpl.nasa.gov/ontology/units.owl#meter_persecond"/>
<swe-om: has DoubleValuerdf: DataType=http://www.w3.org/2001/XML.Schema# Double>6.5</ swe-om:hasDoubleValue>
</swe-om: Quantity>
<swe-om: DataRecordrdf: ID="DataRecord_Atmospheric Conditions">
<swe-om: has Fieldrdf: resource="#Quantity_AirTemperature"/>
```

```
<swe-om:has Fieldrdf:
resource="#Quantity_WindSpeed"/>
<swe: has Definition
rdf:datatype=http://www.w3.org/2001/XM
L.Schema#anyURI>urn:ogc:def:property:
OGC:atmosphericConditions</swe:hasDef
inition>
</swe-om: DataRecord>
```

Figure 1 shows a comparison with SSW and ES3N:

As we can see when we use ES3n, more data are transmitted through network in comparison with SSW. So the lifetime of network decreases more[15].

## VI. CONCLUSION

In recent years progresses in energy efficient design and wireless technologies have enabled various new applications for wireless devices .These applications span a wide range including real time streaming video and audio delivery, remote monitoring using networked micro sensors, personal medical monitoring and home networking of everyday appliances. While these applications require high performance network, they suffer from resource constraints that do not exist in traditional wired computing environments. In particular wireless spectrum is scarce limiting the bandwidth available to applications and making the channel error prone and since the nodes are often battery operated and there is limited available energy. If we can store sensors data more effectively, we have more effective and lifetime sensor networks. In this paper, we compared two methods of sensor data modeling to find better one in some aspect like remaining energy and total data transmission. We should have a tradeoff in choosing sensor data storage method. For future work, we plan to explore a new mechanism to deal with link failures between sensors in the network. Sending data more semantically will also be another step. Another step is evaluating this method when sensors send their data in stream.

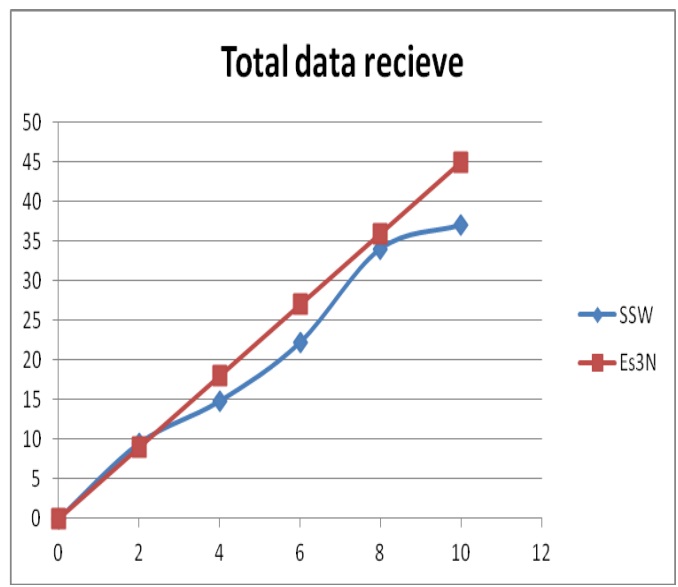

Fig 1. Compares SSW and ES3N. X axis shows number of sensors and Y is the volume of data packets transmitted through network in KB.